\documentclass[twocolumn,showpacs,10pt]{revtex4}
\usepackage{latexsym}
\usepackage{amsmath}
\usepackage{amssymb}
\usepackage{amsthm}

\begin{document}

\title{Multiparticle entanglement and ranks of density matrices}
\author{Bo Chong}
\affiliation{Institut f\"ur Physik, Universit\"at Dortmund, 44221
Dortmund, Germany}
\author{Hellmut Keiter}
\affiliation{Institut f\"ur Physik, Universit\"at Dortmund, 44221
Dortmund, Germany}
\author{Joachim Stolze}
\affiliation{Institut f\"ur Physik, Universit\"at Dortmund, 44221
Dortmund, Germany}
\begin{abstract}
Based on the ranks of reduced density matrices, we derive
necessary conditions for the separability of multiparticle
arbitrary-dimensional mixed states, which are equivalent to
sufficient conditions for entanglement. In a similar way we obtain
necessary conditions for the separability of a given mixed state
with respect to partitions of all particles of the system into
subsets. The special case of pure states is discussed separately.
\end{abstract}
\maketitle

Entanglement is not only at the heart of quantum mechanics but
also a fundamental physical resource in quantum information theory
\cite{Today}. On the one hand, entanglement strongly demonstrates
quantum nonlocality \cite{Nonlocality}, one of the essential
features of quantum mechanics; on the other hand, entanglement
also plays a key role in many parts of quantum communication and
computation \cite{Resource}, such as quantum cryptography
\cite{Cryptography}, quantum dense coding \cite{Densecoding},
quantum teleportation \cite{Teleportation}, quantum one-way
computing \cite{One-way} and so on. Thus it is of great importance
to characterize entanglement of quantum states.

One aspect of characterizing entanglement is to distinguish
entangled states from separable ones. There has been much
significant progress on this problem in many different directions
\cite{Progress,Operational}. Important concepts include Schmidt
decomposition \cite{Schmidt}, Bell inequality \cite{Bell}, partial
transpose \cite{Transpose}, positive maps \cite{Maps}, and
entanglement witnesses \cite{Witness}. However, it is still an
open question how to detect whether a multiparticle
arbitrary-dimensional state is entangled. Another aspect of
characterizing entanglement is to classify entanglement of quantum
states. Based on the separability properties of certain partitions
of systems into subsystems, D\"ur \emph{et al.} \cite{Hierarchy}
have proposed a complete, hierarchic classification of a family of
states, where the states, which have the same number of particles
and the corresponding particles have the same Hilbert-space
dimensions, are put into different levels of a hierarchy with
respect to their entanglement properties. Given a set of many
particles and a partition of that set into parts, it is an
interesting question to ask whether a given multiparticle state is
separable with respect to the given partition. In general, this
question is still open though many partial results
\cite{Hierarchy,Detection} were obtained during the last few
years.

In this work, we propose necessary conditions for the separability
of multiparticle arbitrary-dimensional mixed states based on the
ranks of reduced density matrices. These are equivalent to
sufficient conditions for entangled states. An important advantage
of these conditions is that they are completely operational for
detecting the separability of quantum states. The word
``operational" is used to emphasize, as pointed out by Bru{\ss}
\cite{Operational}, that an operational criterion can be applied
to an explicit density matrix $\rho $, giving some immediate
answer like ``$\rho $ is separable," or ``$\rho $ is entangled,"
or ``this criterion is not strong enough to decide whether $\rho$
is separable or entangled." In a similar way we propose necessary
conditions to determine the separability properties of the
partitions of all particles in a given mixed state, that is, to
determine whether a given mixed state of several subsystems is
entangled. Finally we consider pure states. Two necessary and
sufficient conditions for entangled and fully entangled
\cite{Full} pure states, respectively, are proposed. This allows
us to present a simple procedure to determine the type of
entanglement of a given pure state. In this procedure, we separate
all the particles in a given pure state, without destroying
entanglement of the initial state, into the parts of a special
partition, where every part contains either $X(>1)$ fully
entangled particles or only a single particle.

Let us first consider the definition of entanglement
\cite{Werner}. A pure state $\rho$ of $N$ particles
$A_1,A_2,\cdots,A_N$ is called entangled when it can not be
written as
\begin{equation}
\rho=\rho_{A_1} \otimes \rho_{A_2} \otimes \cdots \otimes
\rho_{A_N} = \bigotimes_{i=1}^{N} \rho_{A_i}
\end{equation}
where $\rho_{A_i}$ is the single-particle reduced density matrix
given by $\rho_{A_i}\equiv \text{Tr}_{\{A_j\}}(\rho)$ for
$\{A_j|\text{all}\ A_j\neq A_i\}$. A mixed state $\rho$ of $N$
particles $A_1,A_2,\cdots,A_N$, described by $M$ probabilities
$p_j$ and $M$ pure states $\rho^j$ as $\rho=\sum_{j=1}^M
p_{_j}\rho^j$, is called entangled when it can not be written as
\begin{equation}
\rho=\sum_{j=1}^M p_j \bigotimes_{i=1}^N \rho^j_{A_i}
\end{equation}
where $p_j > 0$ for $j=1,2,\cdots,M$ with $\sum_{j=1}^M p_j=1$.

For convenience, we will use the following notation. For a state
$\rho$ of $N$ particles $A_1,A_2,\cdots,A_N$, the reduced density
matrix obtained by tracing $\rho$ over particle $A_i$ is written
as $\rho_{R(i)} = \text{Tr}_{A_i}(\rho)$ where $R(i)$ denotes the
set of the remaining $(N-1)$ particles other than particle $A_i$.
In the same way, $\rho_{R(i,j)} =\text{Tr}_{A_j} (\rho_{R(i)}) =
\text{Tr}_{A_j} (\text{Tr} _{A_i}(\rho)) = \text{Tr}_{A_i}
(\text{Tr} _{A_j}(\rho))$ denotes the reduced density matrix
obtained by tracing $\rho$ over particles $A_i$ and $A_j$,
$\rho_{R(i,j,k)} = \text{Tr}_{A_i} (\text{Tr} _{A_j}
(\text{Tr}_{A_k}(\rho))) $, and so on. In view of these relations,
$\rho$ can be called 1-level-higher density matrix of
$\rho_{R(i)}$ and 2-level-higher density matrix of
$\rho_{R(i,j)}$; $\rho_{R(i)}$ can be called 1-level-higher
density matrix of $\rho_{R(i,j)}$ and 2-level-higher density
matrix of $\rho_{R(i,j,k)}$; and so on. It is obvious that the
number of the 1-level-higher density matrices of a reduced density
matrix can be greater than 1. For example, the 1-level-higher
density matrices of $\rho_{R(i,j)}$ are $\rho_{R(i)}$ and
$\rho_{R(j)}$.

The rank of a matrix $\rho$, denoted as $rank(\rho)$, is the
maximal number of linearly independent row vectors (also column
vectors) in the matrix $\rho$. The rank of the density matrix of a
pure state has the following basic property:
\newtheorem{lemma}{Lemma}
\begin{lemma}\label{PureRank}
A state is pure if and only if the rank of its density matrix
$\rho$ is equal to 1, i.e., $rank(\rho)=1$.
\end{lemma}
\begin{proof}
---A state $\rho$ is pure if and only if $\rho^2=\rho$ holds,
that is, $\rho$ is a projection operator onto a one-dimensional
subspace so that only one eigenvalue is equal to 1, all the other
ones being zero. Thus the number of linearly independent row
vectors of $\rho$ is equal to 1. Therefore $rank(\rho)=1$ holds
for a pure state $\rho$.

Conversely, for a density matrix $\rho$ with $rank(\rho)=1$, since
there is only one linearly independent row vector of $\rho$, it is
possible to rewrite the density matrix in a new form with only one
element, whose value is equal to 1, by selecting a suitable basis.
In that basis, $\rho^2=\rho$ is evident and hence $\rho$ is pure.
\end{proof}

Now we discuss necessary conditions for separable states.
\newtheorem*{theorem}{Theorem}
\begin{theorem}[Separability Condition]
If a state $\rho$ of $N$ particles $A_1,A_2,\cdots,A_N$ is
separable, then the rank of any reduced density matrix of $\rho$
must be less than or equal to the ranks of all of its
1-level-higher density matrices, i.e.,
\begin{equation}
rank(\rho_{R(i)})\leq rank(\rho) \label{Criterion}
\end{equation}
holds for any $A_i\in \{A_1,A_2,\cdots,A_N\}$; and
\begin{equation}\label{Criterion2}
\left\{ \begin{split} & rank(\rho_{R(i,j)})\leq rank(\rho_{R(i)}) \\
& rank(\rho_{R(i,j)})\leq rank(\rho_{R(j)})
\end{split} \right.
\end{equation}
holds for any pair of all particles; and so on.
\end{theorem}
\begin{proof}
---Here we will only give the proof for mixed states. Pure states will
be considered in Lemma \ref{Purestate}. For simplicity, we only
prove (\ref{Criterion}). The remaining inequalities can be proved
in a similar way.

A separable mixed state $\rho$ of $N$ particles
$A_1,A_2,\cdots,A_N$ and its reduced density matrix $\rho_{R(i)}$
can be written as
\begin{equation}
\left\{ \begin{split} & \rho=\sum_{j=1}^M p_j \rho^j=\sum_{j=1}^M
p_j \bigotimes_{i=1}^{N} \rho^j_{A_i} \\  & \rho_{R(i)} =
\sum_{j=1}^M p_j\rho^j_{R(i)} = \sum_{j=1}^M p_j\bigotimes_{k=1
\atop k\neq i}^{N} \rho^j_{A_k}.\end{split} \right.
\end{equation}
According to Lemma \ref{PureRank}, any pure state can be
considered a basis vector in its vector space. Thus $M$ pure
states $\rho^j$, where $\rho^j = \bigotimes_{i=1}^{N} \rho^j_{A_i}
\in \bigotimes_{i=1}^{N} \mathcal{H}_{A_i}$ for $j=1,2,\cdots,M$,
are $M$ basis vectors that span a vector space $U \subset
\bigotimes_{i=1}^{N} \mathcal{H}_{A_i}$. Here $\mathcal{H}_{A_i}$
denotes the Hilbert space of particle $A_i$. The maximal number of
linearly independent vectors among these $M$ basis vectors is the
rank of $\rho$, $rank(\rho)$, and at the same time, it is the
dimension of vector space $U$.

In a similar way, $M$ basis vectors $\rho^j_{R(i)}$, where
$\rho^j_{R(i)} = \bigotimes_{\left(k=1 \atop k\neq i\right)}^{N}
\rho^j_{A_k} \in \bigotimes_{\left( k=1 \atop k\neq i \right)}^{N}
\mathcal{H}_{A_k}$ for $j=1,2,\cdots,M$, span a vector space $V
\subset \bigotimes_{\left( k=1 \atop k\neq i \right)}^{N}
\mathcal{H}_{A_k}$ with the dimension $rank(\rho_{R(i)})$.

From the construction of the vector spaces $U$ and $V$ it is clear
that $V$ is a linear subspace of $U$, and hence its dimension is
not greater than that of $U$. This proves (\ref{Criterion}) since
the dimensions of the vector spaces are equal to the ranks of the
density matrices.
\end{proof}

The separability conditions (\ref{Criterion},\ref{Criterion2}) for
mixed states are not sufficient. For example, an important family
of the biqubit mixed states are the so called Werner states
\cite{Werner}, which are mixtures of a maximally entangled biqubit
pure state and the separable biqubit maximally mixed state. These
states are fully characterized by the fidelity $F$, which measures
the overlap of the maximally entangled biqubit pure state with the
Werner states. Though the Werner states do satisfy the
separability conditions (\ref{Criterion},\ref{Criterion2}), they
are entangled for $F>1/2$.

The necessary (but not sufficient) conditions
(\ref{Criterion},\ref{Criterion2}) for a mixed state to be
separable are logically equivalent to the following sufficient
(but not necessary) conditions for a mixed state to be entangled:
\newtheorem{corollary}{Corollary}
\begin{corollary}
Given a mixed state $\rho $, if the rank of at least one of the
reduced density matrices of $\rho$ is greater than the rank of one
of its 1-level-higher density matrices, then the state $\rho$ is
entangled. \label{Entanglement}
\end{corollary}

For a given mixed state, there are hierarchic relations among all
possible partitions of the particles (e.g. in Ref.
\cite{Hierarchy}). For example, consider a partition of all
particles into $i$ parts. If we allow some of the parts to act
together as a new composite part, then we obtain a new partition
into $j$ parts with $j<i$. In a way similar to the proof of the
separability conditions (\ref{Criterion},\ref{Criterion2}), we
obtain the following interesting separability properties of the
partitions of the particles in a given mixed state:
\begin{corollary}\label{Composite}
Consider a mixed state $\rho=\sum_{j=1}^M p_{_j}\rho^j$ and a
partition of the particles. If any two parts $U$ and $V$ in the
partition are separable, that is, the state of the composition
$(U+V)$ of parts $U$ and $V$ can be written as
\begin{equation}
\rho_{(U+V)}=\sum_{j=1}^M p_j\rho^j_{(U+V)}=\sum_{j=1}^M p_j
(\rho^j_U \otimes \rho^j_V)
\end{equation}
where $\rho^j_U \in \mathcal{H}_U$, $\rho^j_V \in \mathcal{H}_V$
and $\rho^j_{(U+V)} \in \mathcal{H}_{(U+V)}$, then the ranks of
the two reduced density matrices $\rho_U$ and $\rho_V$ both are
less than or equal to the rank of $\rho_{(U+V)}$, i.e.,
\begin{equation}
\left\{ \begin{split} & rank(\rho_U)\leq rank(\rho_{(U+V)}) \\
& rank(\rho_V)\leq rank(\rho_{(U+V)}).
\end{split} \right.
\end{equation}
\end{corollary}
The $M$ basis vectors $\rho^j_U$ and the $M$ basis vectors
$\rho^j_V$ span two linear subspaces of the composite vector space
spanned by the $M$ basis vectors $\rho^j_{(U+V)}$. Thus as the
dimensions of the two linear subspaces, $rank(\rho_U)$ and
$rank(\rho_V)$ both are not greater than $rank(\rho_{(U+V)})$, the
dimension of the composite vector space, Corollary \ref{Composite}
is proved. The Werner states again show that the separability
conditions for mixed states in Corollary \ref{Composite} are not
sufficient.

The necessary separability conditions for the partitions in
Corollary \ref{Composite} can again be reformulated as sufficient
entanglement conditions of the partitions: given a mixed state and
a partition of the particles, consider any two parts in the
partition. If the rank of at least one of the reduced density
matrices of the two parts is greater than the rank of the density
matrix of the composition of these two parts, then these two parts
are entangled.

Now we discuss pure states.
\begin{lemma}
A pure state is entangled if and only if the rank of at least one
of its reduced density matrices is greater than 1.
\label{Purestate}
\end{lemma}
\begin{proof}
---If a pure state is entangled, according to Schr\"odinger's
definition of entanglement \cite{Entanglement}: ``The whole is in
a definite state, the parts taken individually are not", then at
least one of the states obtained by tracing the original state
over some particles is mixed. By Lemma \ref{PureRank}, the rank of
this reduced state is greater than 1. Conversely, if the rank of
one reduced density matrix of a pure state is greater than 1, then
the reduced state is mixed, and according to Schr\"odinger's
definition, the original state is entangled.
\end{proof}

An important subclass of the multiparticle entangled states are
the so-called fully entangled states \cite{Full}, which cannot be
reduced to mixtures of states where a smaller number of particles
are entangled. For example, triqubit states that are not of the
forms $\rho_1 \otimes \rho_{23}$, $\rho_2 \otimes \rho_{13}$, and
$\rho_3 \otimes \rho_{12}$, or mixtures of these states are fully
entangled, such as the Greenberger-Horne-Zeilinger (GHZ) state
\cite{GHZ}. In terms of the ranks of reduced density matrices, we
obtain the following necessary and sufficient condition for a pure
state to be fully entangled:
\begin{corollary}\label{Fullyentangled}
A pure state is fully entangled if and only if the ranks of its
all reduced density matrices are greater than 1.
\end{corollary}
\begin{proof}
---A pure state is fully entangled if and only if every particle and
every multi-particle combination in the system are entangled with
the remaining particles. That is, the states of every individual
particle and every individual multi-particle combination are
mixed, i.e., the ranks of all reduced density matrices are greater
than 1, and vice versa.
\end{proof}

For a given pure state $\rho$, if its particles are separated into
two parts $U$ and $V$, then the Schmidt decomposition of state
$\rho$ is written as \cite{Schmidt}
\begin{equation}
\rho = \sum_{i=1}^k \lambda_i |u_i\rangle \langle u_i| \otimes
|v_i\rangle \langle v_i|
\end{equation}
where $|u_i\rangle \in \mathcal{H}_U$, $|v_i\rangle \in
\mathcal{H}_V$ and $\sum_{i=1}^k \lambda_i =1$ with $\lambda
_i>0$. Here the number $k$ is called the Schmidt rank of $\rho$,
which is the rank of the reduced density matrix $\rho_U$ (and
$\rho_V$):
\begin{equation}\label{Bipartiterank}
rank(\rho_U)=rank(\rho_V).
\end{equation}
Then we obtain the following useful Lemma:
\begin{lemma}\label{Partition}
Given a pure state $\rho$, if its particles are separated into two
parts $U$ and $V$, then $rank(\rho_U)=1$ holds if and only if
these two parts are separable, i.e., $\rho = \rho_U \otimes
\rho_V$.
\end{lemma}
\begin{proof}
---If $rank(\rho_U)=1$ holds, then $rank(\rho_V)=1$ holds by
Eq. (\ref{Bipartiterank}), thus states $\rho_U$ and $\rho_V$ are
pure by Lemma \ref{PureRank}. According to the proposition in Ref.
\cite{Correlation}: ``For two systems $U$ and $V$, whenever $U$ is
in a pure state, no correlation exists between $U$ and $V$",
states $\rho_U$ and $\rho_V$ are separable. Therefore the whole
pure state $\rho$ can be written as $\rho=\rho_U \otimes \rho_V$.
Conversely, if $\rho$ is pure and separable with respect to the
two parts $U$ and $V$, that is, $\rho= \rho_U \otimes \rho_V$,
then the ranks obey \cite{Matrix} $rank(\rho) =
rank(\rho_U)*rank(\rho_V)$=1, and hence
$rank(\rho_U)=rank(\rho_V)=1$.
\end{proof}

Using the results obtained above, we construct the following
procedure to find a special partition of a given pure state $\rho$
of $N$ particles $A_1,A_2,\cdots,A_N$, where each part is the
minimal set of particles which cannot be separated any more
without destroying entanglement of the initial state, so that the
particles are separable when they are in different parts but
entangled when they are in one and the same part. Our procedure
consists in successively searching for all subsets of growing size
which are separable from the rest of the system in the sense of
Lemma \ref{Partition}. The maximal set size which has to be
checked for separability is $\lfloor N/2 \rfloor$ (the maximal
integer less than or equal to $N/2$), since along with every
separable set of size $M$, its complement of size ($N-M$) also is
of course separable from all other particles. In more detail the
procedure works as follows:

Step 1. Calculate the rank of $\rho_{R(i)}$ for all particles. By
Lemma \ref{Partition}, if $rank(\rho_{R(i)})=1$ holds, then $\rho$
factorizes as $\rho = \rho_{A_i} \otimes \rho_{R(i)}$. Suppose
there exist $M_1$, $0\leq M_1\leq N$, particles that satisfy
$rank(\rho_{R(i)})=1$, then $\rho$ is the tensor product of $M_1$
single-particle parts and a part of $(N-M_1)$ particles. After
this step, it is impossible that there exists a separable single
particle in the $(N-M_1)$-particle part. If $(N-M_1)>3$
\cite{Reason} holds, we perform the next step, otherwise the
procedure ends.

Step 2. For the part of the remaining $N_2$($=N-M_1$) particles,
calculate the rank of $\rho_{R(i,j)}$ for all two-particle
combinations. If there exist $M_2$, $0\leq M_2 \leq \lfloor N_2/2
\rfloor$, two-particle combinations that satisfy
$rank(\rho_{R(i,j)})=1$, then the part of $N_2$ particles is the
tensor product of $M_2$ two-particle parts and a part of
$(N_2-2M_2)$ particles. If $(N_2-2M_2)>5$ holds, we perform the
next step, otherwise the procedure ends.

The following steps are similar to steps 1 and 2. In the end, if
we obtain separable parts in the procedure, then state $\rho$ can
be written as the tensor product of those parts. If we do not
obtain any separable part in the procedure, then state $\rho$ is
fully entangled.

As an example to explain the procedure in detail, we use the
6-qubit pure state $|\Psi \rangle =(1/2) (|000000\rangle +|000111
\rangle +|011000 \rangle + |011111 \rangle ) $. In step 1, after
calculating $rank(\rho_{R(i)})$ for all qubits, we obtain only
$rank(\rho_{R(1)})=1$ so that $\rho=\rho_{A_1} \otimes
\rho_{R(1)}$. Since $(6-1)>3$, we continue. In step 2, for the
part of the remaining 5 qubits, after calculating
$rank(\rho_{R(i,j)})$ for all 2-qubit combinations, we obtain only
$rank(\rho_{R(2,3)})=1$ so that $\rho_{R(1)} = \rho_{(A_2,A_3)}
\otimes \rho_{R(1,2,3)}$. Since $(5-2)<5$, we end the procedure.
In the end, state $\rho$ can be written as $\rho=\rho_{A_1}
\otimes \rho_{(A_2,A_3)} \otimes \rho_{(A_4,A_5,A_6)}$.

In summary, we have proposed separability criteria for
multiparticle arbitrary-dimensional mixed states in terms of the
ranks of reduced density matrices. Furthermore, we discussed
detection and classification of entanglement in multiparticle pure
states. As compared to the important necessary condition given by
the positivity of the partial transpose \cite{Transpose}, our
results are quite convenient to apply but not quite as strong.
Combinations of the rank and positive partial transpose criteria
have been used to study the separability properties of some
special composite systems\cite{Concrete}. It is an interesting
problem for the further research to investigate the relation
between these two approaches in more detail.

We thank Zeng-Bing Chen for very useful comments on the
manuscript, and Jiangfeng Du, Ping-Xing Chen, De-Cheng Wan, Xinhua
Peng and Jingfu Zhang for valuable discussions.


\begin{thebibliography}{99}
\bibitem{Reason}For $(N-M_1) \leq 3$, the set of all remaining particles
cannot contain further separable subsets since all separable
single particles have already been found by assumption. Similar
considerations hold for the following steps.
\end{thebibliography}
\end{document}